%% file: lks.tex
\documentclass[12pt]{article}
\usepackage{graphicx}
\usepackage{amssymb}
\begin{document}

\newcommand{\Mvariable}[1]{{#1}}
\newcommand{\Mfunction}[1]{{#1}}
\newcommand{\Muserfunction}[1]{{#1}}
\newcommand{\overbar}[1]{{\bar{#1}}}
\newcommand{\imag}{i}

\input{macros}

\begin{titlepage}
\rightline{}

\rightline{hep-th/0408072}

\vskip 2cm
\begin{center}
\Large{{\bf A note on the non-commutative\\ dynamics of spinning D0 branes}}
\end{center}

\vskip 2cm
\begin{center}
Duane Loh\footnote{\texttt{Duane\_Loh@hmc.edu}\ Address after September 2004: Cornell University}\ ,\ Kit Rodolfa\footnote{\texttt{Kit\_Rodolfa@hmc.edu}\ Address after September 2004: Cambridge University} and \ Vatche Sahakian\footnote{\texttt{sahakian@hmc.edu}}
\end{center}
\vskip 12pt
\centerline{\sl Keck Laboratory}
\centerline{\sl Harvey Mudd}
\centerline{\sl Claremont, CA 91711, USA}

\vskip 2cm

\begin{abstract}
Rotational dynamics is known to polarize D0 branes into higher dimensional fuzzy D$p$-branes: the tension forces between D0 branes provide the centripetal acceleration,
and a puffed up spinning configuration stabilizes. In this work, we consider a rotating cylindrical
formation of finite height, wrapping a compact cycle of the background space along the axis of rotation. We find an intriguing 
relation between the angular speed, the geometry of the
cylinder, and the scale of non-commutativity; and we point out a critical radius corresponding to the case where the area of the cylinder is proportional to the number of D0 branes - reminiscent of Matrix black holes. 
\end{abstract}

\end{titlepage}
\newpage
\setcounter{page}{1}

\section{Introduction and results}
\label{intro}

D0 branes are natural probes of Planckian dynamics~\cite{Douglas:1996yp}. In a particularly  interesting regime, D0 brane coordinates are represented by non-commuting matrices; and 
one's intuition stemming from a notion of smooth spacetime quickly falters. Non-commutative dynamics
entails rich new physics, exotic conclusions, and is a testing ground for quantum gravity\cite{Banks:1997hz}-\cite{Chen:2004jz}. It is then
useful to study toy systems involving configurations of D0 branes.

A particularly interesting phenomenon of D0 brane dynamics involves a higher dimensional
D$p$-brane being dynamically weaved out of polarized D0 branes. Typically, background
Ramond-Ramond (RR) fluxes are used to polarize the D0 brane network~\cite{Myers:1999ps,Patil:2004nf}. More recently, Harmark and
Savvidy~\cite{Harmark:2000na} have demonstrated that spinning a spherical lump of D0 branes 
can also lead to a puffed up fuzzy system. In a dual picture, one would talk about a rotating ellipsoidal
D2 brane with magnetic flux on the worldvolume. The basic stabilization mechanism involves 
balancing the tension of strings stretching between the D0 branes with the centripetal
acceleration. In~\cite{Savvidy:2000td}, it was shown that such a system is indeed classically stable. These configurations
can be interesting settings for black brane computations. Similar structures had inspired the modeling 
of Matrix black holes whose thermodynamic properties had been shown to agree with that of black holes in Matrix theory~\cite{Banks:1997hz,Horowitz:1997fr,Banks:1997cm,Banks:1997tn}. Yet another interest in D0 brane fuzz-balls arises in brane-world
discussions. The worldvolume theory of the resulting higher dimensional D$p$-brane
is inherently non-commutative and provides for interesting new phenomena at high
energies, when the scale of non-commutativity is probed~\cite{Seiberg:1999vs}.

In this work, we consider a cylindrical configuration of D0 branes stabilized by spin. The cylinder
is wrapped along its axis of rotation onto a compact cycle of the background space. 
Finite height cylinders are interesting since they allow one to explore the
mixing of several length scales 
in a non-commutative setting: the size of the background compact
cycle or the height of the cylinder, the radius of the cylinder, and the scale of non-commutativity. The size of the compact cycle may also be imagined to be infinite so as to generate an infinite cylinder, as in~\cite{Bak:2001kq}. We would then have a non-commutative worldvolume theory with one non-compact and one compact direction; and
the compact direction is stabilized dynamically through rotation.

\begin{figure}
\centerline {
\includegraphics[height=3in]{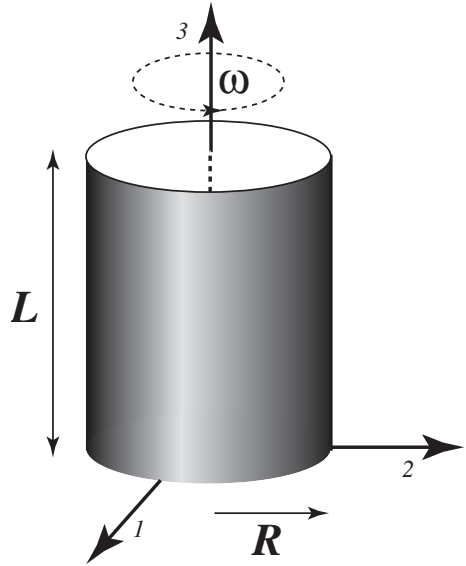}
}
\caption{The coordinates system used to describe the fuzzy cylinder. The direction labeled $3$ is compact of size $L$.}\label{cylinder}
\end{figure}

Our problem is described by the following parameters (see Figure~\ref{cylinder}): $N$ is the number of D0
branes; $L$ is the
size of the compact cycle; $R$ is the radius of the cylinder; $\omega$ is
the angular frequency of rotation;
$w$ and $k_1$ are respectively the winding number along the compact cycle and 
the lateral winding number; finally, the string length $l_s$ sets the scale for lengths, and the string coupling $g_s$
is assumed to be small. To simplify matters, we fix $g_s$, $N$, $w$, and $k_1$ to convenient values, and think of the problem in terms of varying $L$, $R$, and $\omega$. The limit $L\rightarrow\infty$, 
$N\rightarrow \infty$ presumably 
corresponds to the infinite cylinder, which has been studied in the literature in various
settings~\cite{Bak:2001kq,Hashimoto:2004fa,Patil:2004nf,Patil:2004nq}.

To study the dynamics, we use the Dirac-Born-Infeld (DBI) action of~\cite{Tseytlin:1997cs,Myers:1999ps}.
We summarize the highlights of our results whereas the details are left to the main text:

\begin{itemize}
\item We find that the angular speed $\omega$ is related to the other parameters by
\bb\label{eq1}
\omega=-\frac{4\pi w k_1 L}{\lambda N}\ .
\ee
The scale of non-commutativity on the
worldvolume of the cylinder is $4\pi/N$ in string units. Our analysis is valid if
\bb
\Omega\equiv \omega R\ll 1\ \ \ ,\ \ \ \omega l_s\ll 1
\ee
which still leaves $R$ a free parameter. Hence, we now have a two parameter problem.

\item In previous works~\cite{Bak:2001kq,Hashimoto:2004fa}, the case of the
infinite cylinder was studied and 
identified as BPS stable, with the radial extent $R$ being a flat direction free
to be chosen at no cost in energy\footnote{For other work on related configurations see~\cite{Mateos:2001qs,Hyakutake:2003px,Patil:2004nf,Patil:2004nq,Huang:2004uy,Janssen:2003ri,Janssen:2004jz,Ohta1,Ohta2,Ohta3}}. In our work, we adopt a different representation of the algebra that can accommodate a wrapped
cylinder of finite height; with $N$, the number of D0 branes, being finite as well. This involves the mixing of infinite and finite dimensional representations in a delicate recipe.

\item  We argue for the existence of a critical radius $R_{cr}$ given by
\bb
R_{cr}=\frac{1}{\omega}\ ,
\ee
corresponding to the case where the D0 branes are moving with the speed of light. If one
considers
the possibility that the conjectured symmetrized trace prescription 
of~\cite{Tseytlin:1997cs,Myers:1999ps} is valid to all orders in the DBI expansion for the simple matrix
algebra we have\footnote{In general, the validity has been checked to quadratic order in $\lambda$; discrepancies have been identified in general at the third order~\cite{Taylor:1999gq,Taylor:1999pr,Bilal:2001hb,Koerber:2001uu,deBoer:2003cp}. More validation of the possibility that such corrections are not relevant to our analysis may be found in the Discussion section.}, the action vanishes at this critical radius. This is reminiscent of giant graviton dynamics 
arising in a different context~\cite{Brodie:2000yz,Janssen:2003ri,Janssen:2004jz}. Furthermore, at $R_{cr}$, we have
\bb
\mbox{Cylinder area}= \frac{\lambda N}{2}\ .
\ee
With entropy $S\sim N$
for Matrix black holes~\cite{Banks:1997hz,Horowitz:1997fr,Banks:1997tn} - which are
qualitatively similar configurations - this is very suggestive of a holographic statement arising
in a non-commutative theory at weak string
coupling.
\end{itemize}

These conclusions lend themselves to interesting future directions that we discuss in the last section, where we also comment on the issue of classical radiation from the configuration.
In Section 2, we setup the problem and present the solution to the equations of
motion. In Section 3, we discuss the critical radius and the dynamics of perturbation modes. And some details
of the perturbation analysis may be found in Appendix A.

\vspace{0.3in}
{\bf Note Added:} 
In the original version of the preprint, a numerical perturbation analysis was also presented, leading to the identification of unstable modes for certain ratios of the cylinder height to its radius. As pointed out in~\cite{Bak:2001kq}, the algebra describing the cylinder in question saturates a BPS bound, where the central charge is the cylinder's angular momentum, the latter being proportional to the Casimir of the algebra. Hence, the expectation is that the cylinder must be stable in all regions of the parameter space and for all representations of the algebra. In the current version of the preprint, the numerical part of the perturbation analysis has been removed as it apparently leads to unreliable conclusions.

\vspace{0.5in}
\section{The fuzzy rotating cylinder}

We start from the DBI action of~\cite{Tseytlin:1997cs,Myers:1999ps}
\bb\label{DBI}
S_{DBI}=-T_p\int d^{p+1}\sigma\, \mbox{STr}\lk[
e^{-\phi}\sqrt{-\mbox{det}\lk[\mbox{P}\lk[E_
{ab}+E_{ai} (Q^{-1}-\delta)^{ij}E_{jb}\re]+\lambda F_{ab}\re]}\sqrt{\mbox{det} Q^i_j}\re]\ ,
\ee
where
\bb
Q^{i}_j\equiv\delta^i_j+i\lambda\lk[\Phi^i,\Phi^j\re]\ \ \ \ , \ \ \ \lambda\equiv 2\pi l_s^2\ ,
\ee
while the nine $\Phi^i$'s are $N\times N$ hermitian matrices representing
the non-commutative coordinates of $N$ D$p$-branes.
The tension of a D$p$ brane is given by
\bb
T_p=\frac{2\pi}{g_s \lk(2\pi l_s\re)^{p+1}}\ ,
\ee
$\mbox{P}[]$ denotes the pullback to the D-brane worldvolume, and $\mbox{STr}$ stands for the symmetrized trace~\cite{Tseytlin:1997cs,Myers:1999ps}. 
For the task at hand, we consider $N$ D0 branes and zero background fields. In particular, we have
in~\pref{DBI}
\bb
E_{\mu\nu}\equiv G_{\mu\nu}+B_{\mu\nu}\rightarrow\eta_{\mu\nu}\ \ \ ,\ \ \ \phi\rightarrow 0\ ,
\ee
and the Chern-Simons term is vanishing.
This leads to the action
\bb\label{action}
S_{DBI}=-T_0\int dt\ \mbox{STr} \sqrt{1-\lambda^2 {{\dot{\Phi^i}}}^2}\sqrt{\mbox{det} \lk[\delta^i_j+i\lambda \lk[\Phi^i,\Phi^j\re]\re]}
\ee
where we have dropped a term of the form
\bb
\dot{\Phi}^i \dot{\Phi}^j \lk[\Phi_i,\Phi_j\re]\rightarrow 0
\ee
because of the symmetrized trace prescription\footnote{For the configuration we will write, this term is in fact zero irrespective of the symmetrized trace prescription.}, and we have chosen the non-dynamical gauge field on the worldline to be zero
at the expense of imposing the constraint
\bb\label{constraint}
\lk[\dot{\Phi}^i,\Phi_i\re]=0\ .
\ee
If we were interested in an approximate form of the dynamics, we can focus on the regime where\footnote{The
D0 brane coordinates $\Phi^i$ have dimension of inverse length in these units; in particular,
these variables are related to space coordinates by $x\rightarrow \lambda \Phi$.}
\bb\label{cond1}
\dot{\Phi}^i\ll\frac{1}{l_s^2}\ \ \ \ ,\ \ \ \lk[\Phi^i,\Phi^j\re]\ll\frac{1}{l_s^2}\ \ \ ,\ \ \ \ddot{\Phi}^i\ll\frac{1}{l_s^3}\ .
\ee
We then arrive at the simpler expression
\bb\label{approxaction}
S_{DBI}\simeq=-T_0\int dt\ \mbox{STr} \lk[1-\frac{\lambda^2}{2} \lk({D_0{\Phi^i}}\re)^2-\frac{\lambda^2}{4}\lk[\Phi^i,\Phi^j\re]^2+\cdots\re]\ ,
\ee
where the covariant derivative has been reintroduced
\bb\label{theD}
D_0\Phi^i=\dot{\Phi}^i+i\lk[A_0,\Phi^i\re]
\ee
for future reference. For now, we set $A_0=0$ as mentioned above.
The equations of motion are
\bb\label{eom}
\ddot{\Phi}^i=\lk[\Phi^j,\lk[\Phi^i,\Phi_j\re]\re]\ .
\ee
supplemented by the constraint~\pref{constraint}. 

\vspace{0.5in}
\subsection{A time-dependent solution}

Next, our goal is to find a solution to the equations of motion~\pref{eom} subject to the 
constraint~\pref{constraint} representing $N$ D0 branes forming a fuzzy cylinder of finite height and rotating
about its axis of symmetry. We start by singling out three of the space dimensions for
embedding the cylinder. We label these three directions by $\Phi^i$, with $i=1,2,3$. 
Furthermore, we compactify the direction $i=3$ on a circle of circumference $L$ on which 
we intend to wrap the longitudinal extent of the cylinder (see Figure~\ref{cylinder}).
The remaining six transverse polarizations will be denoted by $\Phi^a$. 

Consider a hermitian matrix $\hat{Y}$, and another matrix $\hat{Z}$ with its complex conjugate $\hat{Z}^\dagger$, satisfying 
the closed algebra
\bb\label{algebra}
\lk[\hat{Z},\hat{Z}^\dagger\re]=0\ \ \ ,\ \ \ 
\lk[\hat{Z},\hat{Y}\re]=\omega \hat{Z}\ \ \ ,\ \ \ 
\lk[\hat{Z}^\dagger,\hat{Y}\re]=-\omega \hat{Z}^\dagger\ ,
\ee
where $\omega$ is a real constant.
We will talk about the representation of this algebra in the next section.
We construct two new matrices from $\Phi^1$ and $\Phi^2$
\bb
Z\equiv \Phi^1+i \Phi^2\ \ \ ,\ \ \ Z^\dagger=\Phi^1-i \Phi^2\ .
\ee
And we write 
\bb
Z=z(t) \hat{Z}\ \ \ ,\ \ \ \Phi^3=\frac{w L}{\lambda}\hat{Y}\ \ \ \ ,\ \ \ \Phi^a=0\ .
\ee
We have arranged for the cylinder to wind the $3$ direction $w$ times, with $w$ being
an arbitrary integer.
It is then easy to check that this ansatz satisfies the constraint~\pref{constraint} and
the equations~\pref{eom} provided we have
\bb
z(t)=\frac{R}{\lambda} e^{i\omega t}\ .
\ee
Here, $R$ is the radius of the rotating cylinder as shown in Figure~\ref{cylinder}.

\newpage
\subsection{Algebra representation}

To proceed with analyzing the properties and perturbations of the solution we found in the
previous section, we need to identify a representation of the algebra~\pref{algebra}.
The representation we adopt is worked out in detail in~\cite{Uehara:2004vp}, custom designed to depict
a wrapped fuzzy cylinder. We first introduce two new operators $\rho$
and $\sigma$ 
\bb\label{an}
\hat{Y}=\rho\ \ \ ,\ \ \ \hat{Z}=e^{i k_1 \sigma}\ ,
\ee
satisfying the infinite dimensional algebra.
\bb
\lk[\sigma,\rho\re]=\frac{4\pi i}{N}\ .
\ee
$k_1$ is an integer that appears to represent the number of lateral windings of the cylinder.
Given the compactification in the $\rho$ direction, we may partially truncate this
representation as described in~\cite{Uehara:2004vp}. We focus on $N$ being an odd integer with 
$N=2 M+1$. A basis of $N\times N$ matrices that describes a wrapped non-commutative cylinder is
\bb
e(k')\equiv e^{{k'}_1 \sigma + i {k'}_2 \rho}\ .
\ee
where ${k'}=({k'}_1,{k'}_2)$ represents a pair of integers 
\bb\label{par1}
{k'}_1=p' N+q'\ \ \ \ \mbox{with}\ \ \ q'\in \lk(-M, M\re)\ \ \ \mbox{and}\ \ \ p'\in\lk(-\infty, \infty\re)\ ;
\ee
\bb\label{par2}
{k'}_2=m'\ \ \ \ \mbox{with}\ \ \ m'\in \lk(-M, M\re)\ .
\ee
Note in particular that the modes for ${k'}_2$ have been truncated.
The algebra satisfied by these basis matrices is
\bb
\lk[ e(k'),e(k'')\re]=-2 i \sin \lk(\frac{2\pi}{N} k'\times k''\re) e(k'+k'')\ ,
\ee
where $k'\times k''={k'}_1 {k''}_2-{k'}_2 {k''}_1$. And
\bb\label{alg1}
\lk[\rho, e(k')\re]=\frac{4\pi {k'}_1}{N} e(k')\ .
\ee
The explicit forms are given by~\cite{Uehara:2004vp}
\bbb\label{rep1}
\rho\rightarrow
-2\pi i \del_\theta \lk(
\begin{array}{ccccc}
1 & & & & \\
 & 1 & & & \\
  & & 1 & & \\
  & & & \ddots & \\
  & & & & 1
\end{array}
\re)
+4\pi\lk(
\begin{array}{ccccc}
\frac{M}{N} & & & & \\
 & \frac{M-1}{N} & & & \\
  & & \frac{M-2}{N} & & \\
  & & & \ddots & \\
  & & & & -\frac{M}{N}
\end{array}
\re)\ .
\eee
\bbb\label{rep2}
& &e^{i(p' N+q')\sigma+i m' \rho}\rightarrow \nonumber \\
& &\lk\{
\begin{array}{cl}
\tau^{p'} \gamma^{-m' (q'+1)}
\lk(
\begin{array}{cccccc}
\overbrace{0\ \  \cdots\ \  0}^{q'} & 1 & & & & \\
 & 0 & \gamma^{-2 m'} & & &  \\
 & & \ddots & \ddots & &  \\
& & & & 0 & \gamma^{-2(N-q'-1)m'} \\
 \gamma^{-2(N-q')m'}\tau & & & & & 0  \\
  \ddots & & & & & \vdots \\
  \underbrace{0\ \  \cdots\ \  0}_{q'-1} & \gamma^{-2(N-1)m'}\tau & & & & 0
\end{array}
\re) & q'>0\\
\tau^{p'} \gamma^{-m'}
\lk(
\begin{array}{ccccc}
1 & & & & \\
 & \gamma^{-2 m'} & & & \\
  & & \gamma^{-4 m'} & & \\
  & & & \ddots & \\
  & & & & \gamma^{-2(N-1)m'}
\end{array}
\re) \ \ \ \ \ \ \ \ \ \ \ q'=0& \\
\tau^{p'-1}\gamma^{-m'(q'+1)}
\lk(
\begin{array}{cccccc}
\overbrace{0\ \  \cdots\ \  0}^{N+q'} & 1 & & & & \\
 & 0 & \gamma^{-2 m'} & & &  \\
 & & \ddots & \ddots & &  \\
& & & & 0 & \gamma^{-2(-q'-1)m'} \\
 \gamma^{-2(-q')m'}\tau & & & & & 0  \\
  \ddots & & & & & \vdots \\
  \underbrace{0\ \  \cdots\ \  0}_{N+q'-1} & \gamma^{-2(N-1)m'}\tau & & & & 0
\end{array}
\re) & q'<0
\end{array}
\re.
\eee
Here, we have defined 
\bb
\tau=e^{2\,i \theta}\ \ \ ,\ \ \ \gamma=e^{\frac{2\pi i}{N}}\ ,
\ee
where $\theta$ is a parameter used to label the infinite dimensional sector of the
representation.

Tracing in this representation becomes
\bb
\mbox{Tr}\rightarrow \int d\theta\ \mbox{Tr}\ .
\ee
We then have the orthogonality statement
\bb
\mbox{Tr}\lk[e(k') e(k'')\re]=e^{-\frac{2\pi i}{N} k'\times k''}\mbox{Tr}\lk[e(k'+k'')\re]=
N \delta_{k',-k''}\ ;
\ee
and we also get
\bb\label{zz}
\hat{Z}\hat{Z^\dagger}=\mbox{1}\ \ \ \,\ \ \ {\hat{Z}}^N=\tau^{k_1}\ .
\ee

With these expressions at hand, we may now verify that
\bb
{\Phi^1}^2+{\Phi^2}^2=\frac{R^2}{\lambda^2}\mbox{{\bf 1}}\ .
\ee
It is also easy to check that our solution carries angular
momentum in the 3 direction given by 
\bb
M_{12}\equiv T_0\lambda^2\mbox{Tr}\lk[\dot{\Phi}^2\Phi^1-\dot{\Phi}^1\Phi^2\re]=T_0 N R^2 \omega\ ;
\ee
while the other components are zero. $\omega$ is obviously the angular speed about the $3$ axis. 
More interestingly, we find from~\pref{alg1} that we need 
\bb\label{omega}
\omega=-\frac{4\pi w L k_1}{\lambda N}\ .
\ee
Note that $w$ is the number of windings of the cylinder in the $3$ directions, $k_1$ is the
number of windings in the lateral direction, and $4\pi/N$ is the scale of
non-commutativity on the cylinder's worldvolume in string units. 
Hence, the angular speed is fixed once the 
representation is fixed. We may make the intuitive statement that the angular speed in units of $R$ is
the ratio of the area of the cylinder to the number of D0 branes $N$.

\vspace{0.5in}
\section{A critical radius and perturbations of the cylinder}

Our solution is parameterized by the following constants: $N$, the number of D0 branes; $R$, the radius of the cylinder; $L$, the
compact size of the $3$ direction; $\omega$, the angular speed; and two integers $w$ and $k_1$
representing windings. $\omega$ however was fixed in the previous section (see equation~\pref{omega}). For fixed $N$, $w$ and $k_1$, we are then dealing
with a two parameter problem, $R$ and $L$. 

The solution is valid provided the conditions~\pref{cond1} are satisfied. This translates to
\bb\label{regime}
\Omega=\omega R\ll 1\ \ \ \ ,\ \ \ \omega l_s\ll 1\ .
\ee
Hence, $R$ is unrestricted. We also note that
$\omega\ll 1/l_s$ may be independently achieved by making either $L$ small or $N$ large.

First, we observe that one can now write
\bb
\mbox{det}\lk[\delta^i_j+i\lambda\lk[\Phi^i,\Phi^j\re]\re]=1+\lambda^2 \omega^2 |z|^2\ .
\ee
Substituting our solution in~\pref{action}, the full form of the action, we get
\bb\label{o4}
S_{DBI}=-T_0 N \int dt\ \sqrt{1-\Omega^4}\ .
\ee
However, this assumes that the
symmetrized trace prescription is good for arbitrary order in the expansion of the square
roots in~\pref{action}. While it is possible that, for this particular simple algebra, additional
corrections to the DBI - as well as corrections to the symmetrized trace prescription - may vanish\footnote{More about this issue in the Discussion section.}, this assumption is generally incorrect. For now, our goal however is to use the guess to the extent of
corroborating the existence of a critical maximum radius for which the D0 branes are moving with the speed of light. 
And indeed as confirmed from equation~\pref{o4}, this radius is  
\bb
R_{cr}=\frac{1}{\omega}\ ,
\ee
for which the potential (and action) vanishes. Furthermore, we then have
\bb
\lk(2\pi R_{cr}\re) L=\frac{\lambda N}{2}\ .
\ee
Hence, the area of the cylinder equals $\lambda N/2$, an interesting holographic statement 
as eluded to in the Introduction.

Next, we consider the dynamics of perturbations of the cylinder for $\Omega\ll 1$. We would then write
\bb\label{expansion}
\Phi^1=\Phi^1_0+{\hat{\phi}}^1\ \ \ \ ,\ \ \ 
\Phi^2=\Phi^2_0+{\hat{\phi}}^2\ \ \ \ ,\ \ \ 
\Phi^3=\Phi^3_0+{\hat{\phi}}^3\ \ \ \ ,\ \ \ 
\Phi^a={\hat{\phi}}^a\ \ \ \ ,\ \ \ 
A_0=\hat{a}\ ,
\ee
where the $0$ subscripts refer to the solution of the previous sections. Note that we fluctuate
the gauge field $A_0$ of~\pref{theD} as well. We will then use the equation of motion arising from 
varying $a$ to restrict 
perturbation modes to physical degrees of freedom.
It is more
convenient to parameterize some of these perturbations differently. We write instead\footnote{To see this, first write formally (\ie\ no matrix 
structure necessarily implied)
\bb
\Phi^1=(R+r) \cos (\Theta+\theta)\ \ \ \ ,\ \ \ 
\Phi^2=(R+r) \sin (\Theta+\theta)\ .
\ee
Then, expand for small $r$ and $\epsilon$ and elevate things back to matrices while symmetrizing over
ambiguous orderings.
We then relabel things in terms of dimensionless parameters $\theta\rightarrow\hat{\epsilon}_\theta$
and $r\rightarrow\hat{\epsilon}_r$.}
\bb
{\hat{\phi}}^1=-(\Phi^2_0 \hat{\epsilon}_\theta+\hat{\epsilon}_\theta \Phi^2_0)+(\Phi^1_0 \hat{\epsilon}_r+\hat{\epsilon}_r \Phi^1_0)\ .
\ee
\bb
{\hat{\phi}}^2=(\Phi^1_0 \hat{\epsilon}_\theta+\hat{\epsilon}_\theta \Phi^1_0)+(\Phi^2_0 \hat{\epsilon}_r+\hat{\epsilon}_r \Phi^2_0)\ .
\ee
Now, we expand the
small perturbations in the matrix basis of the previous section
\bb
\hat{\epsilon}_\theta=\sum_{k'}\theta_{k'} e(k')\ \ \ \ ,\ \ \ \ 
\hat{\epsilon}_r=\sum_{k'}r_{k'} e(k')\ .
\ee
\bb
{\hat{\phi}}^3=\sum_{k'} \frac{\epsilon^3_{k'}}{R} e(k')\ \ \ \ ,\ \ \ \ {\hat{\phi}}^a=\sum_{k'} \frac{\epsilon^a_{k'}}{R} e(k')\ \ \ \ ,\ \ \ \ \hat{a}=\sum_{k'} {a_{k'}} e(k')\ .
\ee
Note that we have made sure that $\epsilon^3_{k'}$ and $\epsilon^a_{k'}$ are
dimensionless for convenience; and we have reality conditions relating modes of opposite signs, such as
$r_{k'}^*=r_{-k'}$. We also 
use the shorthand (see equations~\pref{par1} and~\pref{par2})
\bb
\sum_{k'}\equiv \sum_{p'=-\infty}^{\infty}\sum_{q'=-M}^{M}\sum_{m'=-M}^{M}\ ,
\ee
To assure that an expansion in the $\varepsilon$'s makes sense, we require
\bb\label{conditions}
r_{k'} , \theta_{k'}\ll 1\ \ \ ,\ \ \ {{\varepsilon}}^3_{k'}\ll \frac{N\omega R}{4\pi k_1}\ .
\ee
Substituting~\pref{expansion} into~\pref{approxaction}, and using the commutation relations
\bb
\lk[\Phi^3_0, e(k')\re]=\frac{w L}{\lambda} \frac{4\pi {k'}_1}{N} e(k')\ ;
\ee
\bb
\lk[Z_0,e(k')\re]=-2 i \frac{R}{\lambda} e^{i\omega t} \sin\lk(\frac{2\pi}{N} k\times k'\re) e(k+k')\ ;
\ee
\bb
\lk[Z^\dagger_0,e(k')\re]=2 i \frac{R}{\lambda} e^{-i\omega t} \sin\lk(\frac{2\pi}{N} k\times k'\re) e(-k+k')\ ,
\ee
one can write the worldvolume theory of small perturbations. We emphasize that this analysis is restricted to the regime prescribed by~\pref{regime}.

As far as perturbation modes in the six transverse directions are concerned, we find that all these modes are diagonal with mass squared given by
\bb\label{transverse}
m_{trans}^2=
{\frac{4 R^2}{\lambda^2}
          {\sin^2 \lk(\frac{2\,\pi \,{k_1}\,{l_2}}{N}\re)}
           +{\omega }^2\,
          \frac{{{l_1}}^2}{k_1^2}} >0\ ,
\ee
where $l_1$ and $l_2$ are integers to be described below.

The interesting physics occurs in the 1-2-3 subspace where there are two physical
degrees of freedom (due to the constraint in the system). The details of this analysis are collected in Appendix A. The final action is given by $S_1+S_2+S_3$, where $S_1$, $S_2$, and $S_3$ are defined in~\pref{s1}, \pref{s2}, and \pref{s3} respectively.

\vspace{0.5in}
\section{Discussion}

We end with a few comments about the matter of classical radiation. In the dual picture, the system is to be described by a cylindrical 
D2 brane with magnetic and electric fields on the worldvolume accounting for the dissolved D0 branes 
and rotation. 
It is easy to see that, from this static viewpoint, one expects zero radiation from the RR 1-form
gauge field. 
In the picture involving the D0 branes explicitly, zero 1-form radiation is a much more non-trivial
statement. Yet, it is expected that the conclusion agrees with the dual picture, presumably the two being
related by some Seiberg-Witten map~\cite{Seiberg:1999vs}. Considering the Chern-Simons coupling from~\cite{Myers:1999ps},
one easily finds
\bb
C_i= \frac{J_i(k,\omega)}{{\vec{k}}^2-\omega^2}\ ,
\ee
where $C_i(k,\omega)$ is the Fourier mode of the 1-form RR gauge field. 
The source is identified as~\cite{Myers:1999ps}
\bb\label{}
J_i(k,\omega)\simeq \int dt\ e^{-i\omega t} \mbox{STr} \lk[e^{-i\lambda \Phi^j k_j} {\dot{\Phi}}^i\re]\ .
\ee
Considering linear order in $\lambda$, it is easy to check that 
$\mbox{STr} [\Phi^j {\dot{\Phi}}^i]$ is time independent resulting in zero radiated power.
Indeed, one can see that this pattern repeats to all orders: we have $\dot{Z}=\omega Z$
and ${\dot{Z}}^\dagger=-\omega Z^\dagger$; and hence terms of the form
$\mbox{STr } x\cdots x$ where $x$ stands for a $Z$ or a $Z^\dagger$ (wavefronts cannot
have a component in the 3 direction by symmetry). The number of 
$Z$'s and $Z^\dagger$'s must equal so that the trace does not vanish (see in particular equation~\pref{zz}). But this also
cancels all time dependences from the source leaving zero radiated power through the
RR 1-form. This suggests that the symmetrized trace prescription and the DBI action seem to be working well
for our solution to all orders in $\lambda$. It would also be interesting to see what conclusions are reached with
regards to gravitational and RR 3-form radiation.

\vspace{0.5in}
\section{Appendix A: Some details of the perturbation analysis}

In this appendix, we collect some of the details of the perturbation analysis. 
For notational convenience, we relabel some of the
perturbations as $\theta_{k'}\equiv \epsilon_{k'}^1$ and $r_{k'}\equiv \epsilon_{k'}^2$.
Expanding our action to quadratic order (for 
$\epsilon_{k'}^1 , \epsilon_{k'}^2\ll 1$, ${{\epsilon}}^3_{k'}\ll {N\omega R}/{4\pi k_1}$) in the small perturbations, one finds three
pieces 
after a bit of algebra: from the kinetic term without the gauge field fluctuations 
taken into account, we get
\bbb\label{s1}
S_1&=&N\,R^2\,{\omega }^2 + 4\,N\,R^2\,{\omega }^2\,{\epsilon }_0^2 + 
  4\,N\,R^2\,\omega \,{{\dot{\epsilon}
         }_0}^1 \nonumber \\ &+& 4\,N\,R^2\,{\cos^2 (\frac{2\,\pi \,k\times l}
       {N})}\,\left( {{\dot{\epsilon}}^1_{-l}}+\omega \,{\epsilon}_{-l}^2 
      \right)
     \,\left({{\dot{\epsilon}}^1_l}+ \omega \,{\epsilon}_{l}^2
      \right)  \nonumber \\ &+&
   4\,N\,R^2\,{\cos^2 \lk(\frac{2\,\pi \,k\times l}{N}\re)}\,
   \left( {{\dot{\epsilon}}^2_{-l}}-\omega \,{\epsilon}_{-l}^1 
      \right)
     \,\left({{\dot{\epsilon}}^2_l} -\omega \,{\epsilon}_{l}^1
      \right)  \nonumber \\ &+&
   N\,{\lambda }^2\,{{\dot{\epsilon}}^3_
      {-l}}\,{{\dot{\epsilon}}^3_l}
   + N\,{\lambda }^2\,{{{\dot{\epsilon}}^a
         }_{-l}}\,{{\dot{\epsilon}}^a_
      l}\ . 
\eee
From the quartic potential, we obtain
\bbb\label{s2}
S_2&=&-N\,R^2\,{\omega }^2 - 
  4\,N\,R^2\,{\omega }^2\,{\epsilon }_0^2 - 
  {4\,N\,R^2\,{\omega }^2\,
     {\cos^2 \lk(\frac{2\,\pi \,k\times l}{N}\re)}\,
     {\left( 1 + \frac{{{l_1}}^2}{{{k_1}}^2} \right)} \,{\epsilon}_{-l}^1\,
     {\epsilon}_{l}^1} \nonumber \\ &-& 
  { 8\,\imag  \,N\,R^2\,{\omega }^2\,
     {\cos^2 \lk(\frac{2\,\pi \,k\times l}{N}\re)}\,\frac{l_1}{k_1}\,
     {\epsilon}_{-l}^2\,{\epsilon}_{l}^1}  + 
     {8\,\imag \,N\,R^2\,{\omega }^2\,
     {\cos^2 \lk(\frac{2\,\pi \,k\times l}{N}\re)}\,\frac{l_1}{k_1}\,
     {\epsilon}_{-l}^1\,{\epsilon}_{l}^2} \nonumber \\
     &+& 
  {4\,N\,R^2\,{\cos^2 \lk(\frac{2\,\pi \,k\times l}{N}\re)}\,
     \left( -2\,\frac{R^2}{\lambda^2} - {\omega }^2 + 
          2\,\frac{R^2}{\lambda^2}\,\cos \lk(\frac{4\,\pi \,k\times l}{N}\re)\,
         - \,{\omega }^2\,\frac{{{l_1}}^2}{k_1^2} \right) \,
     {\epsilon}_{-l}^2\,{\epsilon}_{l}^2} \nonumber \\ &-& 
  {2\,N\,R^2\,\omega \,\sin (\frac{4\,\pi \,k\times l}
       {N})\,\frac{{{l}}_1}{k_1}\,{\epsilon}_{-l}^3\,{\epsilon}_{l}^1}
        - {2\,N\,R^2\,\omega \,
     \sin \lk(\frac{4\,\pi \,k\times l}{N}\re)\,\frac{{{l}}_1}{k_1}\,
     {\epsilon}_{-l}^1\,{\epsilon}_{l}^3}\nonumber \\  &+&
         4\,\imag  \,N\,R^2\,\omega \,
   \sin \lk(\frac{4\,\pi \,k\times l}{N}\re)\,{\epsilon}_{-l}^3\,
   {\epsilon}_{l}^2 - 
  4\,\imag  \,N\,R^2\,\omega \,
   \sin \lk(\frac{4\,\pi \,k\times l}{N}\re)\,{\epsilon}_{-l}^2\,
   {\epsilon}_{l}^3 \nonumber \\ &-& 4\,N\,R^2\,
   {\sin^2 \lk(\frac{2\,\pi \,k\times l}{N}\re)}\,
   {\epsilon}_{-l}^3\,{\epsilon}_{l}^3 - 
  {N\,\left( 4\,R^2\,{\sin^2 \lk(\frac{2\,\pi \,k\times l}{N}\re)}
          + {\lambda }^2\,{\omega }^2\,\frac{{{l_1}}^2}{k_1^2} \right) \,
     {\epsilon }_{-l}^a\,{\epsilon }_l^a}
\ .
\eee
Finally, we solve the equations of motion for the gauge field fluctuation modes $a_{l}$
\bbb
a_l&=&-\frac{2\, \imag \,{k_1^2}\, R^2}{4\,R^2\,{\sin^2 \lk(\frac{2\,\pi \,k\times l}{N}\re)}\,
     {{k_1}}^2 + {\lambda }^2\,{\omega }^2\,{{l_1}}^2} \nonumber \\
     &\times & \left( 2\,\omega \,
       {\cos^2 \lk(\frac{2\,\pi \,k\times l}{N}\re)}\,
       {\epsilon}_{l}^1 -\sin \lk(\frac{4\,\pi \,k\times l}{N}\re)\, \lk({{\dot{\epsilon} }_l}^1+
       2\,\,\omega \,
       {\epsilon}_{l}^2 
       \re) +
      {\lambda }^2\,\frac{\omega}{2\,R^2} \,\frac{l_1}{k_1}\,
       {{\dot{\epsilon}}^3_l} \right) 
\eee
and substitute back into the action; this gives the additional piece
\bbb\label{s3}
S_3&=&-\frac{N\,R^4}{4\,R^2\,{\sin^2 \lk(\frac{2\,\pi \,k\times l}{N}\re)}\,
     {{k_1}}^2 + {\lambda }^2\,{\omega }^2\,{{l_1}}^2} \nonumber \\
     &\times & \,\left(
      \frac{{\lambda }^2}{R^2}\,\omega \,{l_1}\,
       {{\dot{\epsilon}}^3_{-l}}+4\,\omega \,
       {\cos^2 \lk(\frac{2\,\pi \,k\times l}{N}\re)}\,{k_1}\,
       {\epsilon}_{-l}^1 -2\,{{{k}}_1}\,\sin \lk(\frac{4\,\pi \,k\times l}{N}\re)\,\lk(
      {{ \dot{\epsilon} }_{-l}^1}+2\, \omega \,
       {\epsilon}_{-l}^2
        \re)
      \right) \nonumber \\ 
     &\times &\left(
      \frac{{\lambda }^2}{R^2}\,\omega \,{l_1}\,
       {{\dot{\epsilon}}^3_l} +4\,\omega \,
       {\cos^2 \lk(\frac{2\,\pi \,k\times l}{N}\re)}\,{k_1}\,
       {\epsilon}_{l}^1 -2\,{{{k}}_1}\,\sin \lk(\frac{4\,\pi \,k\times l}{N}\re)\,\lk({{\dot{\epsilon}}_l^1}+2\,\omega \,
       {\epsilon}_{l}^2
       \re)\right) 
\eee
Our system is now given by $S_1+S_2+S_3$. 
%\bibliographystyle{hplain}
%\bibliography{biblio}

\end{document}

%% file: macros.tex
%%%%%%%%%%%%%%%%%%%%%%%%%%
% Vatche Sahakian's macros

\newcommand{\bb}{\begin{equation}}
\newcommand{\ee}{\end{equation}}
\newcommand{\bbb}{\begin{eqnarray}}
\newcommand{\eee}{\end{eqnarray}}
\newcommand{\diag}{\mbox{diag }}
\newcommand{\Str}{\mbox{STr }}
\newcommand{\Tr}{\mbox{Tr }}
\newcommand{\Det}{\mbox{Det }}
\newcommand{\C}[2]{{\lk [{#1},{#2}\re ]}}
\newcommand{\AC}[2]{{\lk \{{#1},{#2}\re \}}}
\newcommand{\kk}{\hspace{.5em}}
\newcommand{\vc}[1]{\mbox{$\vec{{\bf #1}}$}}
\newcommand{\mc}[1]{\mathcal{#1}}
\newcommand{\del}{\partial}
\newcommand{\lk}{\left}
\newcommand{\ave}[1]{\mbox{$\langle{#1}\rangle$}}
\newcommand{\re}{\right}
\newcommand{\pd}[1]{\frac{\del}{\del #1}}
\newcommand{\pdd}[2]{\frac{\del^2}{\del #1 \del #2}}
\newcommand{\Dd}[1]{\frac{d}{d #1}}
\newcommand{\sech}{\mbox{sech}}
\newcommand{\pref}[1]{(\ref{#1})}

\newcommand
{\sect}[1]{\vspace{20pt}{\LARGE}\noindent
{\bf #1:}}
\newcommand
{\subsect}[1]{\vspace{20pt}\hspace*{10pt}{\Large{$\bullet$}}\mbox{ }
{\bf #1}}
\newcommand
{\subsubsect}[1]{\hspace*{20pt}{\large{$\bullet$}}\mbox{ }
{\bf #1}}

\def\ie{{\it i.e.}}
\def\eg{{\it e.g.}}
\def\cf{{\it c.f.}}
\def\etal{{\it et.al.}}
\def\etc{{\it etc.}}

\def\e{{\mbox{{\bf e}}}}
\def\AA{{\cal A}}
\def\BB{{\cal B}}
\def\CC{{\cal C}}
\def\DD{{\cal D}}
\def\EE{{\cal E}}
\def\FF{{\cal F}}
\def\GG{{\cal G}}
\def\HH{{\cal H}}
\def\II{{\cal I}}
\def\JJ{{\cal J}}
\def\KK{{\cal K}}
\def\LL{{\cal L}}
\def\MM{{\cal M}}
\def\NN{{\cal N}}
\def\OO{{\cal O}}
\def\PP{{\cal P}}
\def\QQ{{\cal Q}}
\def\RR{{\cal R}}
\def\SS{{\cal S}}
\def\TT{{\cal T}}
\def\UU{{\cal U}}
\def\VV{{\cal V}}
\def\WW{{\cal W}}
\def\XX{{\cal X}}
\def\YY{{\cal Y}}
\def\ZZ{{\cal Z}}

\def\sinh{{\rm sinh}}
\def\cosh{{\rm cosh}}
\def\tanh{{\rm tanh}}
\def\sgn{{\rm sgn}}
\def\det{{\rm det}}
\def\trace{{\rm Tr}}
\def\exp{{\rm exp}}
\def\sh{{\rm sh}}
\def\ch{{\rm ch}}

\def\ell{{\it l}}
\def\str{{\it str}}
\def\lp{\ell_{{\rm pl}}}
\def\blp{\overline{\ell}_{{\rm pl}}}
\def\ls{\ell_{{\str}}}
\def\bls{{\bar\ell}_{{\str}}}
\def\bM{{\overline{\rm M}}}
\def\gs{g_\str}
\def\gym{{g_{Y}}}
\def\geff{g_{\rm eff}}
\def\eff{{\rm eff}}
\def\r11{R_{11}}
\def\kel{{2\kappa_{11}^2}}
\def\kten{{2\kappa_{10}^2}}
\def\lpten{{\lp^{(10)}}}
\def\alp{{\alpha '}}
\def\alpe{{{\alpha_e}}}
\def\le{{{l}_e}}
\def\aleff{{\alp_{eff}}}
\def\sqaleff{{\alp_{eff}^2}}
\def\tgs{{\tilde{g}_s}}
\def\talp{{{\tilde{\alpha}}'}}
\def\tlp{{\tilde{\ell}_{{\rm pl}}}}
\def\tr11{{\tilde{R}_{11}}}
\def\wtilde{\widetilde}
\def\what{\widehat}
\def\hlp{{\hat{\ell}_{{\rm pl}}}}
\def\hr11{{\hat{R}_{11}}}
\def\hf{{\textstyle\frac12}}
\def\coeff#1#2{{\textstyle{#1\over#2}}}
\def\CY{Calabi-Yau}
\def\lessapprox{\;{\buildrel{<}\over{\scriptstyle\sim}}\;}
\def\greaterapprox{\;{\buildrel{>}\over{\scriptstyle\sim}}\;}
\def\inbar{\,\vrule height1.5ex width.4pt depth0pt}
\def\IC{\relax\hbox{$\inbar\kern-.3em{\rm C}$}}
\def\IR{\relax{\rm I\kern-.18em R}}
\def\IP{\relax{\rm I\kern-.18em P}}
\def\Z{{\bf Z}}
\def\R{{\bf R}}
\def\One{{1\hskip -3pt {\rm l}}}
\def\sst{\scriptscriptstyle}
\def\osc{{\rm\sst osc}}
\def\lam{\lambda}
\def\lc{{\sst LC}}
\def\pr{{\sst \rm pr}}
\def\cl{{\sst \rm cl}}
\def\D{{\sst D}}
\def\bh{{\sst BH}}
\def\vev#1{\langle#1\rangle}